\documentclass[aps,prstab,twocolumn,groupedaddress,
showpacs,amsmath,amssymb,floatfix,letterpaper]{revtex4}

\usepackage{graphicx}
\usepackage[usenames]{color}

\begin{document}

\title{One 4-Twist Helix Snake to Maintain Polarization in 8-120\,GeV Proton Rings}

\author{F.~Antoulinakis}
\author{E.A.~Ljungman}
\author{A.~Tai}
\author{C.A.~Aidala}
\author{E.D.~Courant}\altaffiliation[also ]{BNL}
\author{A.D.~Krisch}
\author{W.~Lorenzon}
\author{P.D.~Myers}
\author{R.S.~Raymond}
\author{D.W.~Sivers}\altaffiliation[also ]{Portland Physics Institute}
\affiliation{Spin Physics Center, University of Michigan,
Ann Arbor, Michigan 48109-1040, USA}

\author{M.A.~Leonova}
\affiliation{Fermi National Accelerator Laboratory,
Batavia, Illinois 60510, USA}

\author{Y.S.~Derbenev}
\author{V.S.~Morozov}
\affiliation{Thomas Jefferson National Accelerator Facility,
Newport News, Virginia 23606, USA}

\author{A.M.~Kondratenko}
\affiliation{Science and Technique Lab "Zaryad"
630090, Akademika Lavrentieva prospect 6/1, off. 33, Novosibirsk, Russia}

\date{September 4, 2013}

\begin{abstract}
Solenoid Siberian snakes have successfully maintained polarization in particle rings below 1\,GeV,
but never in multi-GeV rings because the Lorentz contraction of a solenoid's
$\int \! B\cdot dl$ would require impractically long high-field solenoids.
High energy rings, such as Brookhaven's 255\,GeV Relativistic Heavy Ion Collider (RHIC),
use only odd multiples of pairs of transverse B-field Siberian snakes directly opposite each other.
When it became impractical to use a pair of Siberian Snakes in Fermilab's 120\,GeV Main Injector (see Fig. 2),
we searched for a new type of single Siberian snake, which should overcome all depolarizing
resonances in the 8.9--120\,GeV range.
We found that one snake made of one 4-twist helix and 2 short dipoles could maintain the polarization.
This snake design might also  be used at other rings, such as Japan's
30\,GeV J-PARC, the 12--24\,GeV NICA proton-deuteron collider at JINR-Dubna,
and perhaps RHIC's injector, the 25\,GeV AGS.

\end{abstract}

\maketitle

To study the strong interaction's spin dependence with polarized proton
beams~\cite{dgcrabb,spin00,spin02,spin04,spin06,spin08},
one must preserve and control the polarization during acceleration and storage
This can be difficult due to many depolarizing (spin) resonances.
In 1977 Derbenev and Kondratenko~\cite{Derbenev:1977} proposed a  clever way to
overcome all depolarizing resonances with arrangements of magnets now called Siberian snakes.
Each snake rotates each beam particle's spin by 180$^\circ$.
This idea was first publicized in the West by Courant~\cite{courant} who coined the name Siberian snake.
The concept was first tested at the Indiana University Cyclotron Facility (IUCF) Cooler ring in
1989~\cite{adk} using a single longitudinal solenoid snake, which is very efficient at 100\,MeV energies,
but inefficient above a few GeV. In a ring with one snake, the protons' stable spin direction is in the horizontal
plane; and it is longitudinal exactly opposite the snake. In a ring with snake pairs exactly opposite each other, the stable spin direction is vertical.
The Derbenev and Kondratenko papers~\cite{Derbenev:1977,Derbenev:1978} noted that for transverse
magnetic fields the $\int \! B\cdot dl$ seen by protons is almost invariant under the Lorentz transformation above a few GeV. This is because the $\int \! B\cdot dl$ needed for a 180$^\circ$ spin rotation only increases like  the velocity,  $\beta$.

More recently two other facilities have used solenoid Siberian snakes to achieve polarized electron beams,
the Amsterdam Pulse Stretcher~\cite{Amps} and MIT's Bates South Hall Ring~\cite{mit}.
Others commissioned studies of how to accelerate and maintain polarized proton beams
including,  SSC~\cite{ssc}, Fermilab's Tevatron~\cite{fermi}, and DESY's HERA~\cite{Alekseeva:1996tt}.
Moreover, Brookhaven has fabricated and used two helical 4\,T Siberian snakes in each RHIC ring  to
successfully accelerate, store and collide 100--255\,GeV polarized protons~\cite{Alekseev:2003sk}.

\begin{figure*}[b]
\includegraphics[width=1.9\columnwidth]
{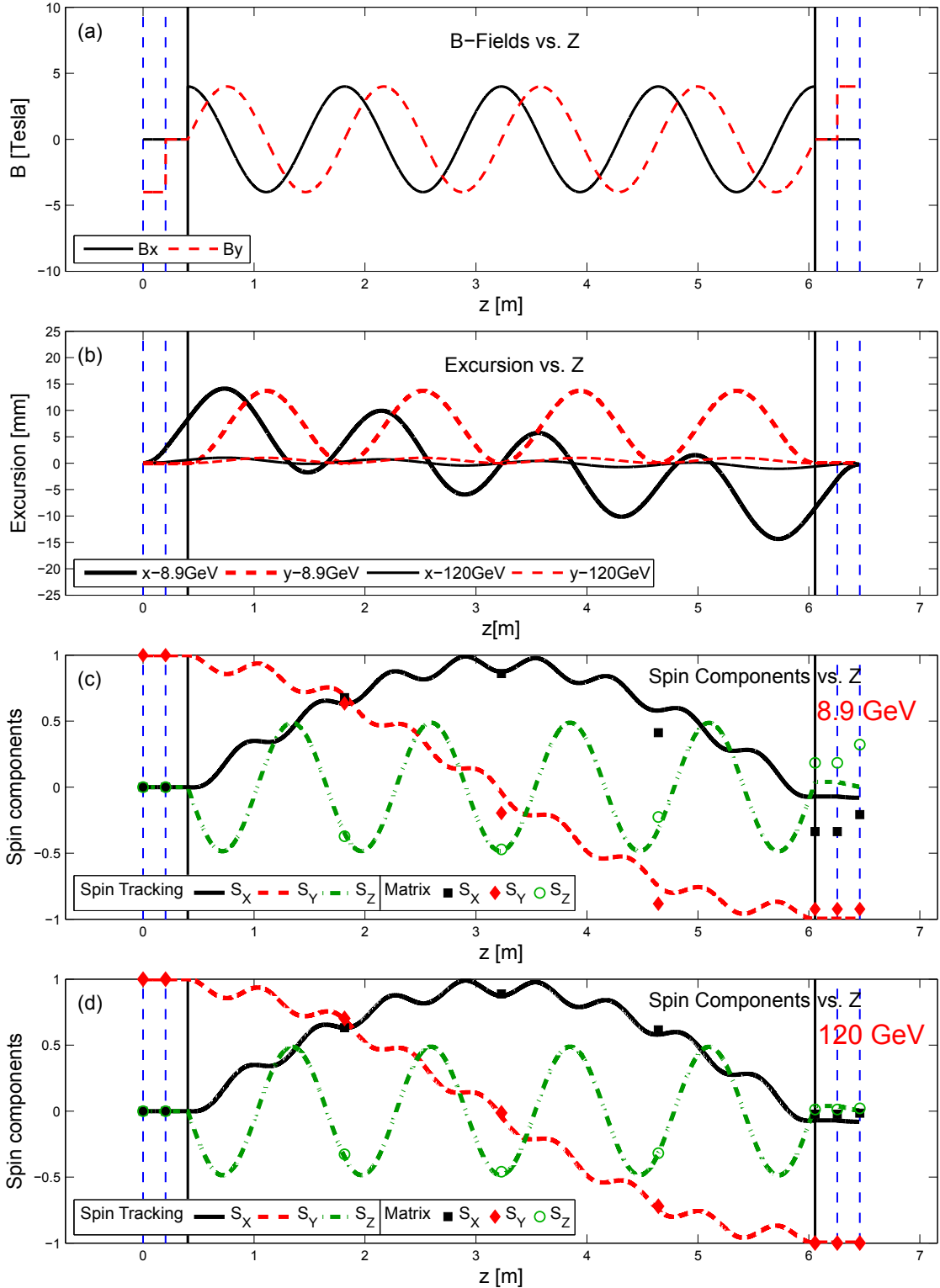}
\vspace{-0.15in}
\caption{
Graphs showing properties of Siberian Snake containing: a 4\,T, 4-twist  5.653\,m helical dipole, a 4\,T, 0.203\,m dipole at each end with 0.200\,m gaps. Vertical solid black lines show helix's edges.
Dashed blue lines show dipoles' edges. The dashed and solid curves are from a Python-based spin tracking program.
The symbols are for analytic matrix calculations, which are more precise, but give results only at edges of each magnet.
(a) Horizontal and vertical $B$-field components as function of $z$.
(b) Horizontal and vertical orbit excursions as function of $z$.
(c) Radial, vertical, and longitudinal spin components as function of $z$ for 8.9\,GeV beam.
(d) Spin components for 120\,GeV beam. (Table I shows detailed numbers)}.
\label{fig:trackingCalcs}
\end{figure*}

\begin{figure*}
  \centering
  \includegraphics[scale=.65,angle=0]{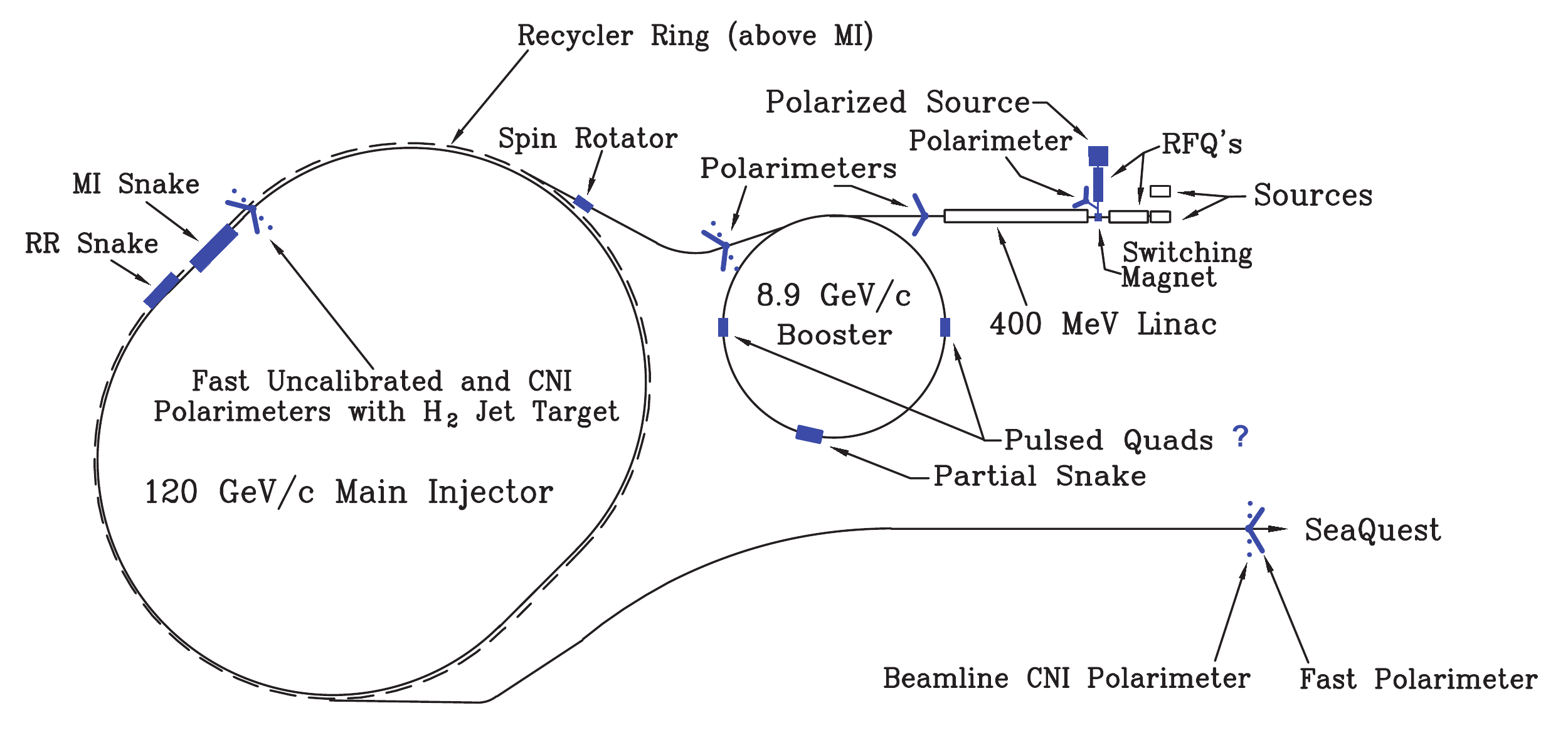}
  \caption{
Main Injector accelerator complex conceptual layout showing  equipment needed for polarized proton beam (in blue).
}\label{fig:polAccelComplex}
\end{figure*}

In flat horizontal rings, a beam proton's spin precesses around the ring dipole magnets' vertical fields.
The spin tune $\nu_s = G\:\gamma$ is the number of spin precessions during one turn around the ring,
where $\gamma$ is a proton's Lorentz energy factor and $G = (g - 2)/2 = 1.792\:847$ is its gyromagnetic
anomaly.
Horizontal magnetic fields can perturb the proton's stable polarization creating depolarizing spin
resonances~\cite{stora,khoe,montague,khiari,sylee}, which occur whenever
\vspace{-0.05in}
\begin{equation}
\nu_s = l\nu_x + m\nu_y + n,
\end{equation}
where $l$, $m$ and $n$ are integers;
$\nu_x$ and $\nu_y$ are the horizontal and vertical betatron tunes, respectively.
Imperfection spin resonances occur when $l~=~m~=~0$.
Intrinsic spin resonances occur when either $l \neq 0$ or $m \neq 0$, or both;
the sum $|l|+|m|$ defines each resonance's order.

A proper snake must rotate the vertical spin component by exactly $\pi$ (180$^\circ$). It must also be optically transparent, which means that, to the rest of the ring, it appears to be an empty space.
Moreover, the excursions caused by the snake must fit well inside the beam's vacuum chamber.
To make a helical snake optically transparent requires a pair of short dipoles with opposite vertical B-fields
upstream and downstream of the helix.

At Fermilab's 120\,GeV Main Injector (see Fig.~2), the beam excursions caused by a single helix were too large at its 8.9\,GeV injection energy. Thus, we studied the properties of a multi-twist helix, which was first discussed by Kondratenko~\cite{kondra82} and then studied in detail by Courant~\cite{courant88},
who considered only helical snakes with twist angles of $3\pi , 5\pi , 7\pi$ , etc.
To obtain optical transparency, these need both a horizontal and a vertical dipole at each end.
Thus, we focused on multi-twist snakes that need only one vertical dipole at each end, with
twist angles of $2\pi , 4\pi , 6\pi$, $etc.$; and finally chose $8\pi$.

A 1-twist helical dipole's magnetic field  is given by
\vspace{-0.05in}
\begin{equation}
\begin{array}{c}
B_x = B_0 \sin{kz},   \
B_y = B_0 \cos{kz},
\end{array}
\end{equation}
where the coordinates are: $x$ (radial), $y$ (vertical) and $z$ (longitudinal);
while $k = 2 \pi / L$, and $L$ is the helix's length.
The magnetic fields, the protons' resulting excursions and spin behavior in the 4-twist helix were obtained by treating it as four 1-twist helices in series. The beam excursions were obtained by inserting into the tracking program the Lorentz force,
\vspace{-0.05in}
\begin{equation}
\begin{array}{c}
\vec{F} = q\vec{v}\times\vec{B}.\
\end{array}
\end{equation}
Numeric spin tracking calculations were made using the magnetic fields from Eq. (2) in 780,000 steps along
the snake's 6.459\,m length.
At each point along the snake axis, the change in spin was calculated using the Thomas-BMT equation~\cite{lht,bmt},
\vspace{-0.05in}
\begin{equation}
\begin{array}{c}
\frac{d\vec{S}} {dt} = -\frac{q} {\gamma m} [(1 + G\gamma)\vec{B_\perp} + (1 + G)\vec{B_\parallel}]\times\vec{S}.\
\end{array}
\end{equation}
The resulting spin components are shown in Fig. 1,
along with the magnetic field components and the horizontal and vertical excursions.

Due to concern that fringe fields might reduce the polarization, spin tracking calculations were also performed assuming a fringe field which drops linearly from 4\,T at 1,2,3,4 or 5\,cm inside to 2\,T at each edge of each magnet and to 0\,T at 1,2,3,4 or 5\,cm past each edge. As shown in Table I, these unrealistically large changes in the fringe field were found to have no effect on the precisely 180$^\circ$ spin flip of the vertical spin component; they did have small effects on the radial and longitudinal components, but a snake's ability to maintain a stable polarization at the point opposite the snake is affected only by its ability to flip the vertical polarization by 180$^\circ$.

Analytic excursion calculations were also done using the equation for a 1-twist helix obtained by Courant~\cite{courant96}
\vspace{-0.05in}
\begin{equation}
\small{\begin{array}{c}
x(z) = x_0 +(x'_0-kr_0\cos{ks_0})(s-s_0)+r_0(\sin{ks}-\sin{ks_0})\\
y(z) = y_0 +(y'_0-kr_0\sin{ks_0})(s-s_0)-r_0(\cos{ks}-\cos{ks_0}),
\end{array}}
\end{equation}
where $r_0=\frac{1}{(k^2\rho)}$.
The spin calculations  were done using the matrix  obtained by Syphers ~\cite{syp} and a new matrix for linearly changing fields.
Each 1-twist helical dipole rotates the spin by angle $\mu$ around the helix's spin rotation axis $\vec{n}$ according to
\vspace{-0.05in}
\begin{equation}
\begin{array}{c}
\mu = - 2 \pi \sqrt{1 + (\frac{\kappa}{k})^2} \\
\vec{n} = \frac{\hat{z}+(\frac{\kappa}{k})\hat{y}}{\sqrt{1+(\frac{\kappa}{k})^2}},
\end{array}
\end{equation}
where $\kappa = (1+G\gamma )\frac{B_0}{B \rho}$, and where the magnetic rigidity, ($B \rho$) is the ratio of a proton's momentum to its charge.
The matrix for  the 4-twist helix was obtained by multiplying four 1-twist helices in series.

Note that, in going from a 1-twist to 4-twist helix, the maximum transverse excursions inside the helical snake decrease almost 4-fold (from 4.76\,mm to 1.26\,mm at 120\,GeV and 64.18\,mm to 16.55\,mm at 8.9\,GeV), while the total snake length increases by only 29\% (from 4.242\,m to 5.459\,m).

These analytic matrix calculations were done for fringe fields from the dipoles but no fringe fields from the 4-twist helix, since we found a matrix for the dipole fringe fields by taking the derivative of Eq. 4 and solving the resulting linear differential equation, but we were unable to find a matrix for the 4-twist helix's fringe fields.
For the 180$^\circ$ vertical spin flip these matrix calculations are consistent with the corresponding spin tracking calculations which includes both dipole and helix fringe fields at the $10^{-3}$ level at  120\,GeV and deviate at the 0.06 level at 8.9\,GeV, as shown in Fig.~1 and Table~I. It also shows that the spin tracking calculations are independent of the fringe field length at the $2~10^{-6}$ level  at 120\,GeV and at the $10^{-4}$ level at 8.9\,GeV. Moreover, all 5 analytic matrix calculations are independent of the dipole fringe field length, (confirmed by these calculations at the $10^{-8}$ level at both 120\,GeV and  8.9\,GeV). Thus, the effect of any fringe fields should be negligible.

We found that by changing the beam energy by a few GeV from the nominal 120\,GeV, the polarization direction at the end of the beam-line can be rotated within the horizontal plane or into the vertical direction as required by the Fermilab E-1027, the Polarized Drell-Yan experiment~\cite{Drell-Yan exp}. This is  at the end of a 2.4\,km beam with ~140 bending magnets, some with horizonal magnetic fields, some with vertical fields and others with skew fields.

\begin{table}
\caption{Numerical values of the radial (x), vertical (y) and longitudinal (z) spin components as the (Fringe) field length is increased from 0\,cm to $\pm 5$\,cm for 8.9 and 120\,GeV.}
\centering
\begin{tabular*}{\columnwidth}{ c c c c c c c }
 \hline
 \hline
    8.9\,GeV & \multicolumn{2}{c}{X spin} & \multicolumn{2}{c}{Y spin component} & \multicolumn{2}{c}{Z spin} \\
  \hline
        Fringe & M & T & Matrix & Track & M & T \\
  \hline
  0 (cm) & -0.208 & -0.080 & -0.9234 & -0.9967 & 0.3223 & 0.00083 \\
  $\pm 1$\,cm & -0.208 & -0.081 & -0.9234 & -0.9967 & 0.3223 & 0.00085 \\
  $\pm 2$\,cm & -0.208 & -0.081 & -0.9234 & -0.9967 & 0.3223 & 0.00096 \\
  $\pm 3$\,cm & -0.208 & -0.081 & -0.9234 & -0.9967 & 0.3223 & 0.00101 \\
  $\pm 4$\,cm & -0.208 & -0.080 & -0.9234 & -0.9968 & 0.3223 & 0.00137 \\
  $\pm 5$\,cm & -0.208 & -0.080 & -0.9234 & -0.9968 & 0.3223 & 0.00164 \\

 \hline
 \hline
\end{tabular*}
\label{Table Ia}
\end{table}

\begin{table}
\centering
\begin{tabular*}{\columnwidth}{ c c c c c c c }
 \hline
 \hline
       120\,GeV & \multicolumn{2}{c}{X spin} & \multicolumn{2}{c}{Y spin component} & \multicolumn{2}{c}{Z spin} \\
  \hline
        Fringe & M & T & Matrix & Track & M & T \\
  \hline
  0\,cm & -0.018 & -0.0004 & -0.9995 & -1.000 & 0.0254 & 0.00000 \\
  $\pm 1$\,cm & -0.018 & -0.0004 & -0.9995 & -1.000 & 0.0254 & 0.00000 \\
  $\pm 2$\,cm & -0.018 & -0.0004 & -0.9995 & -1.000 & 0.0254 & 0.00000 \\
  $\pm 3$\,cm & -0.018 & -0.0002 & -0.9995 & -1.000 & 0.0254 & 0.00000 \\
  $\pm 4$\,cm & -0.018 & -0.0000 & -0.9995 & -1.000 & 0.0254 & -0.00001 \\
  $\pm 5$\,cm & -0.018 &  0.0004 & -0.9995 & -1.000 & 0.0254 & -0.00003 \\

 \hline
 \hline
\end{tabular*}
\label{Table Ib}
\end{table}

In summary, by using two independent spin calculation techniques, we showed that one can flip the vertical spin component by very near $\pi$ (180$^\circ$)  by using a new type of single Siberian snake with transverse magnetic fields. This should maintain the polarization for proton beams in the Fermilab Main Injector up to 120\,GeV.  This single snake containing a 5.653\,m, 4-twist, 4\,T single helical dipole, with a 0.203\,m, 4\,T dipole on either end should maintain the proton polarization in the Fermilab Main Injector up to 120\,GeV. Such a transverse field 4-twist single-snake configuration opens new possibilities for polarized proton beams in rings where it is difficult to place pairs of snakes at points 180$^\circ$ apart.

We thank I. Korbanis and H. B. White Jr. for their advice on Fermilab's evolving Main Injector.
We thank the University of Michigan for supporting this work.

\end{document}